# AODVSEC: A Novel Approach to Secure Ad hoc On-demand Distance Vector (AODV) Routing Protocol from Insider Attacks in MANETs

Akshai Aggarwal[1], Dr. Savita Gandhi[2], Nirbhay Chaubey[2], Pathik Shah[2], Madhvi Sadhwani[2]

[1]Gujarat Technological University, Ahmedabad – 380 015, Gujarat, India
`vc@gtu.ac.in`

[2]Department of Computer Science, Gujarat University, A'bad – 380 009, Gujarat, India
`drsavitagandhi@gmail.com, nirbhay@ieee.org,`
`pathikmshah88@gmail.com, madhvi.sadhwani@gmail.com`

## ABSTRACT

*Mobile Ad hoc Network (MANET) is a collection of mobile nodes that can communicate with each other using multihop wireless links without requiring any fixed based-station infrastructure and centralized management. Each node in the network acts as both a host and a router. In such scenario, designing of an efficient, reliable and secure routing protocol has been a major challenging issue over the last many years. Numerous schemes have been proposed for secure routing protocols and most of the research work has so far focused on providing security for routing using cryptography. In this paper, we propose a novel approach to secure Ad hoc On-demand Distance Vector (AODV) routing protocol from the insider attacks launched through active forging of its Route Reply (RREP) control message. AODV routing protocol does not have any security provision that makes it less reliable in publicly open ad hoc network. To deal with the concerned security attacks, we have proposed AODV Security Extension (AODVSEC) which enhances the scope of AODV for the security provision. We have compared AODVSEC with AODV and Secure AODV (SAODV) in normal situation as well as in presence of the three concerned attacks viz. Resource Consumption (RC) attack, Route Disturb (RD) attack, Route Invasion (RI) attack and Blackhole (BH) attack. To evaluate the performances, we have considered Packet Delivery Fraction (PDF), Average End-to-End Delay (AED), Average Throughput (AT), Normalized Routing Load (NRL) and Average Jitter and Accumulated Average Processing Time.*

## 1. INTRODUCTION

Mobile Ad hoc Network (MANET) is a collection of mobile nodes which establish a network spontaneously and communicate over a shared wireless channel without any pre-deployed infrastructure and mostly decentralized management [1]. In an ad hoc network, security is became a key requirement now a days. As the communication is done via shared wireless media, it is highly vulnerable to many security threats. In this particular case, security can be thought for two occurrences viz. Data communication and Routing message relays. Securing data communication is very much easy using point-to-point and/or end-to-end security techniques, but the major concerns are resided within the security of routing message as these messages ultimately form a route to pass the data packets between two ends. Thus, routing protocol itself has to be secured and none of the basic protocols are given with security provision.

In this paper, we have demonstrated Active Forge exploits that are possible against the AODV protocol and have offered a secure solution using a simple cache concept to provide security from the malevolent nodes.

The rest of this paper is organized as follows: Section 2 describes the fundamental working of AODV. Section 3 elaborates the concerned security issues in AODV. Our related study is kept in Section 4. Section 5 provides the complete understanding of our proposed approach to modify AODV protocol. Section 6 and Section 7 show our Methodology for evaluation and results of our Simulation in NS-2.33 along with its Graphs respectively. We have concluded this paper in Section 8 followed by the vital references for our research work.

## 2. FUNDAMENTAL WORKING OF AODV

Ad hoc On Demand Distance Vector Routing (AODV) [2]-[3] is a Reactive routing protocols, which appear to be more suitable for ad hoc networks, do not maintain up-to-date information about the network topology as it is done by the proactive routing protocols like Destination Sequenced Distance Vector (DSDV) [4] and Optimized Link State Routing Protocol (OLSR) [5], but they create routes on demand. Among reactive routing protocols, the Ad hoc On Demand Distance Vector Routing (AODV) is a very popular on demand routing protocol which has been developed for MANETs and it is widely use. AODV performance is the best considering its ability to maintain connection by periodic exchange of information and we found that AODV outperforms other routing protocol and is viable for MANET [6] - [8].

As its name suggests, route discovery process is initiated on demand i.e. whenever a particular source node wants to start communication with a particular destination and if it is not having the corresponding route in its routing table, it initiates route discovery process. Route discovery process consists of two control messages viz. Route Request (RREQ) and Route Reply (RREP). The source node generates RREQ message at the very first step of the discovery process and floods that to its neighborhood. Intermediate node(s) upon having the RREQ message, makes reverse route entry for the source node in its routing table and floods the same RREQ packet further in the network.

The ultimate destination node will generate RREP packet in response to the received RREQ message and sends that back to the originator of RREQ via unicast property. Before passing the received RREP packet towards the source node, intermediate node(s) again updates its routing table now for having the destination node's forwarding entry into it. Thus, Source node can start the actual data transmissions after receiving RREP for the concerned destination. Figure 1 depicts the fundamental working of route discovery process in AODV.

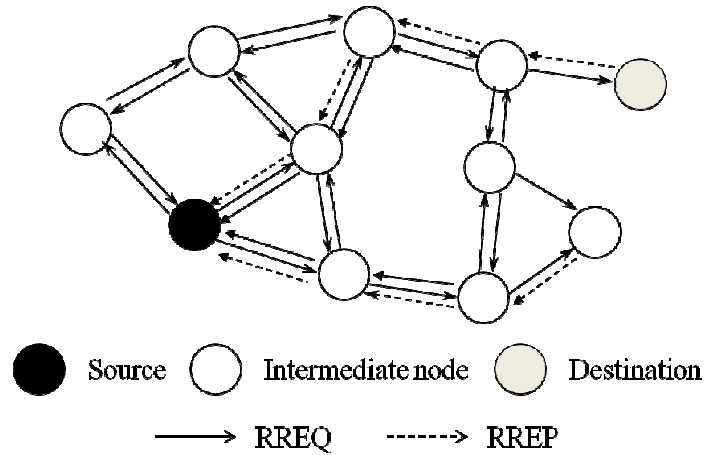

Figure 1. Fundamental Working of Route Discovery Process in AODV

One of the very important properties of AODV is that it also allows any intermediate node(s) to generate and send RREP towards the source node on behalf of the ultimate destination node if it is having the fresher route to that destination node. Besides the mentioned two control messages, AODV uses one more to maintain the established route as a part of route maintenance process in AODV using Route Error (RERR) message.

## 3. SECURITY ISSUES WITH AODV

In any kind of communications, security is considered as an essential requirement for making the system much more reliable. For the same purpose, routing protocol in ad hoc networks is the most important aspect that has to be secured from possible security threats. Widely used AODV routing protocol is also highly vulnerable to many attacks as there is no security provision with it and the control messages are dispatched in plain text only. Thus, attacks on AODV are mainly made by forging any of its control messages viz. RREQ, RREP, RERR. [9].

In this work, we focused on the three perilous active forge attacks through fake RREPs in which attacker actively makes use of malignant RREP message by forging the fields of destination sequence number and/or hop count to launch the attacks on already established path in-between known pair of source and destination. Following subsections explain the effect of the concerned actively forged attacks by fake RREP and blackhole attacks.

### 3.1. Resource Consumption (RC) Attack

For achieving this particular goal, an attacker uses two fake RREPs and further disturbs the route as if it makes a loop in the data path resulting into resource consumption of the participating nodes of data path. Figure 2(a) shows one example topology in which a path is already established.

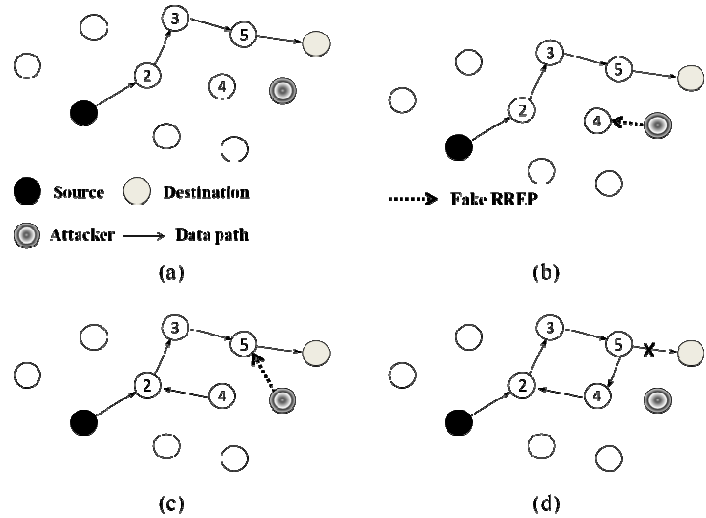

Figure 2. Loop forming attack using two fake RREPs

Figure 2(b) and Figure 2(c) illustrates the fake RREP propagation by the attacker impersonating node 2 and node 4 respectively. As a result, the victim nodes inside the data loop would be forced to consume more energy in continuous data forwarding (see Figure 2(d)).

### 3.2. Route Disturb (RD) Attack

The major goal of doing such an attack is to break or disturb the established route between known pair of source and destination. Figure 3 shows the consequence of this attack. Above Figure 2 (a) is repeated in Figure 3 (a) for showing the example of an established route and the position of an attacker.

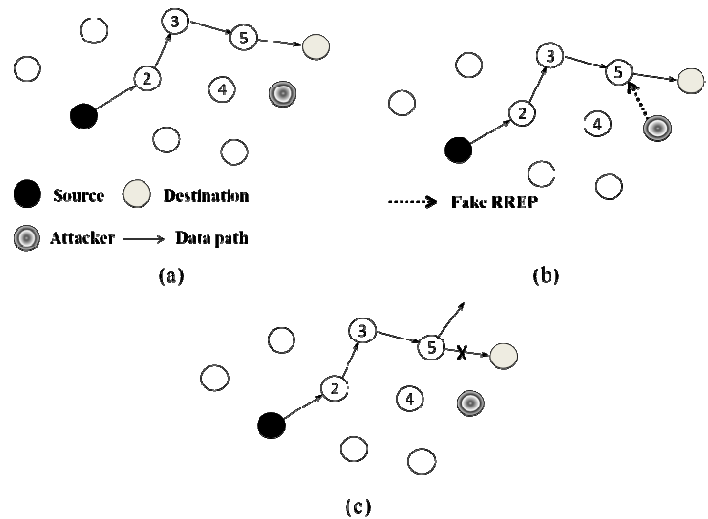

Figure 3. Route Disturb attack using a fake RREP

After finding a route to node-5, as shown in Figure 3 (b) the attacker generates and sends a fake RREP to node-5 impersonating a non-existing node identity forcing node-5 to change its routing table entry for the destination. As a result, the existing route is disturbed (see Figure 3 (c)) and new route discovery process would be expected.

### 3.3. Route Invasion (RI) Attack

By means of such an attack, the attacker node invades itself into the existing data path for having all data packets pass through it. To achieve this particular goal, the attacker will first find a route to reach the victim pair of source and destination. Figure 4 demonstrates the route invasion attack. Figure 2 (a) is repeated for Figure 4 (a) showing the position of attacker for this case.

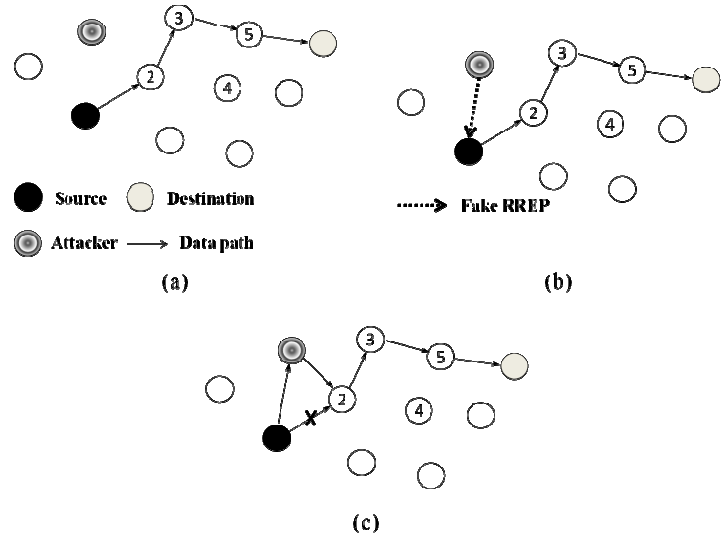

Figure 4. Route Invasion attack using a fake RREP

Figure 4 (b) shows that the attacker node generates and sends a fake RREP to the source node forcing it to change its routing entry for reaching the destination via itself. Figure 4 (c) depicts the resultant data path after successful launch of this particular attack.

### 3.4. Blackhole (BH) Attack

In blackhole attack [13]-[14], the malicious node waits for the neighbors to initiate a RREQ packet. As the node receives the RREQ packet, it immediately sends a false RREP packet with a modified higher sequence number. So, that the source node assumes that node is having the fresh route towards the destination. The source node ignores the RREP packet received from other nodes and begins to send the data packets over malicious node. A malicious node takes all the routes towards itself. It does not allow forwarding any packet anywhere. This attack is called a blackhole as it swallows all data packets.

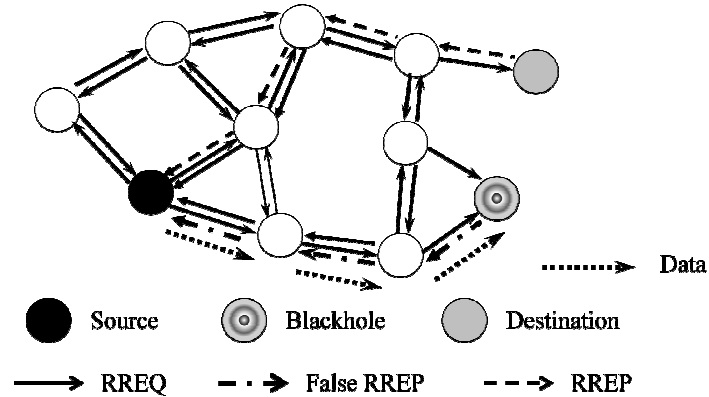

Figure 5. Blackhole attack

Figure 5 shows the Blackhole attack in MANETs in which, source node wants to send data packets to a destination node in the network. The attacker replies with false reply RREP having higher modified sequence number. So, data communication initiates from the source towards the blackhole instead of the destination.

## 4. RELATED WORK

AODV Routing protocol is vulnerable to a wide range of security threats; insider and outsider attacks [9]. So, to provide the security and prevent from such kind of attacks many authors have proposed numerous techniques as Manel Guerrero Zapata [10] proposed SAODV secure routing protocol, which uses digital signatures to authenticate non-mutable fields and hash chains to authenticate the hop-count field in both RREQ and RREP messages. Dahill et al. [11] proposed ARAN, a routing protocol for ad hoc networks that uses authentication and requires the use of a trusted certificate server. In ARAN, every node that forwards a route discovery or a route reply message must also sign it, (which is very computing power consuming and causes the size of the routing messages to increase at each hop). Naldurg, and R. Kravets [12] proposed the Security –Aware Ad hoc Routing (SAR) protocol which defines the trust degree that should be associated with each node, and ensures that a node is prevented from handling a RREQ (Route Request) unless it provides the required level. This way, data packets will be sent only through trusted nodes, with respect to the defined level. All the above mechanisms provide the security to AODV using either cryptography or centralized trusted server which consumes the more computational time. Our proposed security extension to AODV (AODVSEC) gives the similar level of security but with reduced computational time.

## 5. INTRODUCTION TO AODVSEC

In the basic AODV, there is no reason with any node to reject fake RREPs although it has not witnessed the concerned route discovery process for which incoming RREP is given to it. Upon receiving that particular message (though it is fake), a node will always check its routing table for probable updates. AODVSEC is introduced to secure the protocol from the above explained active forge attacks commenced through fake RREP packets. We propose to maintain a cache containing the important information to validate any incoming RREP packet. As the entries inside this cache are created during the fundamental route discovery process i.e. in response to the generated RREQ message, we call it as a Route REQest ACKnowledgement Cache (RREQ-ACK Cache).

## 5.1. Applicability Statement

AODVSEC is proposed to secure AODV protocol from the insider attacks launched through actively forged RREP messages only. The attacker generates the fake RREP by means of increasing the destination sequence number and/ or decreasing the hop count to forcefully update the victim node's routing table. AODVSEC is not considering the attacks based on the fake RREQ message. Moreover, our research work focus especially on network layer hence the impersonation would only be done by means of spoofing the IP addresses and not the MAC addresses.

## 5.2. RREQ-ACK Cache

The main purpose for defining cache is to validate every incoming RREP packet. Those specific fields which help a node to uniquely identify the expected RREP in response to the witnessed RREQ message for the specific pair of source and destination were desired to be in cache. Therefore, we have used the immutable unique fields of route discovery process viz. Destination IP Address and RREQ Timestamp to copy into at the specific position in the cache. Thus, this information has to be collected from all those neighbors along with their identity at which the current RREQ message was accepted and further processed.

We have used the re-flooded message by the neighbor to populate the Cache. Thus, before dropping the duplicated message, a node would be having its cache entry at the time of route discovery process itself. For this, an extra field is introduced in the basic message format of AODV which helps a node for taking decision whether to make cache entry from each received duplicated RREQ packet. Moreover, we have also suggested a new control message named Route REQest ACKnowledgement (RREQ-ACK) (see the next subsection) when RREQ re-flooding is not available, like RREP generation. Figure 6 provides a vision to the proposed cache.

| RREQ-ACK Cache | | | | |
|---|---|---|---|---|
| Nb | D | T | F | Ex |
|  |  |  |  |  |
|  |  |  |  |  |

Figure 6. RREQ-ACK Cache

Where, **Nb** = Neighbor node address, from which either concerned duplicated RREQ or RREQ-ACK is received

**D** = Destination node address, directly copied from the same received message for which route discovery process is witnessed

**T** = Timestamp, again directly copied from the same received message for which route discovery process is witnessed

**F** = Boolean Flag, indicates the activity behind the making of this entry. If 0, entry is made using the duplicated RREQ and if 1, entry is made using the RREQ-ACK

**Ex** = Expiry Time, after this timeout this entry has to be flushed; the value of expiry time is decided based on the following equation:

$$Ex = CURRENT\_TIME + PATH\_DISCOVERY\_TIME \qquad (1)$$

## 5.3. Control Messages of AODVSEC

### 5.3.1. Route Request (RREQ)

| Type=<br>RREQ | J | R | G | D | U | A | Reserved | Hop<br>Count |
|---|---|---|---|---|---|---|---|---|
| RREQ Broadcast ID ||||||||||
| Destination IP Address ||||||||||
| Destination Sequence Number ||||||||||
| Originator IP Address ||||||||||
| Originator Sequence Number ||||||||||
| **RREQ Timestamp** ||||||||||
| **Previous Node IP Address** ||||||||||

Figure 7. Modified Message Format of RREQ

**A** = Acknowledgement Flag, if set then AODVSEC procedures have to be followed

**RREQ Timestamp** = Time at which the route discovery process is initiated

**Previous Node IP Address** = Node address from which node has received RREQ. It is updated before re-flooding. Initially, it is same as Originator IP Address. Whenever any node receives the duplicated RREQ, it compares its identity with the value from this field and upon matching does the cache entry before dropping the packet.

### 5.3.2. Route Request ACKnowledgement (RREQ-ACK)

The node(s) which will send RREP instead of further re-flooding of RREQ have to first generate and send it back to the node from which it has received RREQ message.

| Type=<br>RREQ-ACK | Reserved |
|---|---|
| Own IP Address ||
| Destination IP Address ||
| RREQ Timestamp ||

Figure 8. Proposed Message Format for RREQ-ACK

**Type** = RREQ-ACK, indentifies the packet type

**Own IP Address** = IP address of message generating node

**Destination IP Address** = directly copied from the received RREQ

**RREQ Timestamp** = directly copied from the received RREQ

### 5.3.3. Route Reply (RREP)

| Type= RREP | R | A | Reserved | Prefix Size | Hop Count |
|---|---|---|---|---|---|
| Destination IP Address ||||||
| Destination Sequence Number ||||||
| Originator IP Address ||||||
| Lifetime ||||||
| RREQ Timestamp ||||||

Figure 9. Modified Message Format of RREP

**RREQ Timestamp** = directly copied from the received RREQ

### 5.4. Validation of RREP using RREQ-ACK Cache

It is very simple process for deciding whether incoming RREP message is to be accepted or not. For doing the same, one-to-one mapping is done with the information resided in the RREQ-ACK Cache.

As shown in Figure 10, Destination IP Address from incoming RREP is matched with D field in the RREQ-ACK Cache. Likewise, RREQ Timestamp from RREP is matched with T field in the Cache.

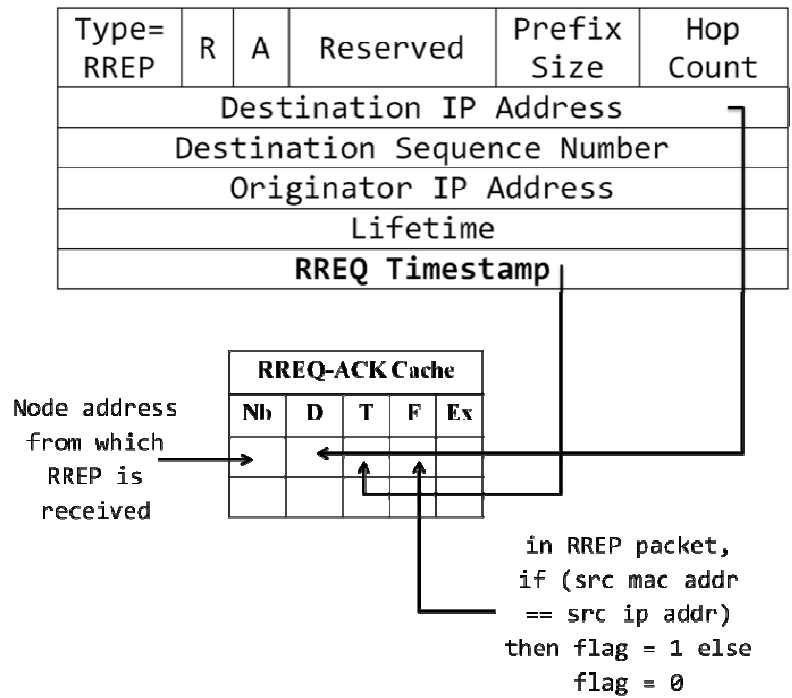

Figure 10. Validation of received RREP using RREQ-ACK

If all matches are exactly found same (i.e. Destination IP address from RREP with 'D' field of cache, RREQ Timestamp from RREP with the 'T' field of cache and the Sending node address with the 'nb' field of cache, only then the received RREP is accepted and further process is done in the same way as in AODV.

## 5.5. Incorporation with the fundamental working of basic AODV

To implement our proposed algorithm we modify the basic functionality of AODV as follows:

- **Upon receiving the RREQ (recvRequest function):**

    - If RREQ is duplicated then, node is supposed to compare its identity with "Previous Node IP Address" field in RREQ. Upon success, the concerned information is stored in the RREQ-ACK cache followed by the normal processing.

    - If RREP is to be generated then, RREQ-ACK is sent back before sending the RREP to the same node.

- **Upon receiving RREQ-ACK (recvRequestAck function – newly defined):** Information from the message is stored in the RREQ-ACK Cache.

- **Upon receiving RREP (recvReply function):** Validation through RREQ-ACK Cache. Upon success, acceptance of RREP followed by the normal processing.

## 5.6. Fundamental route discovery process of AODVSEC

Figure 11 shows the fundamental route discovery process of AODVSEC. It is almost similar to Figure 1 of basic AODV as far as control message propagation in the network is concerned.

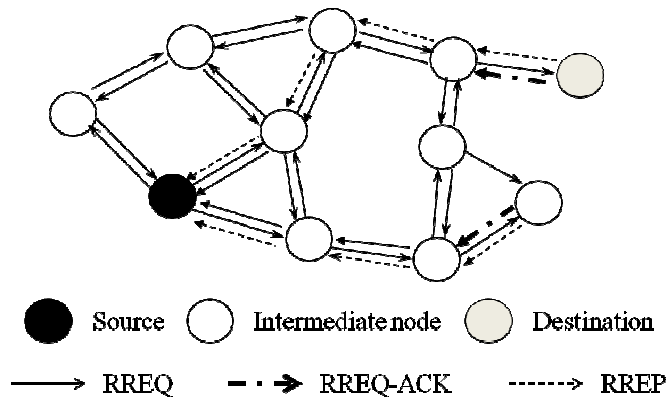

Figure 11. Route Discovery Process of AODVSEC

It can be seen that the nodes which generated RREP, had sent RREQ-ACK messages before doing the same. The RREQ-ACK Cache is created in back ground to this process just like routing table is handled along with the process. Whenever any RREP message is received at any node, it would be first validated using the previously maintained information in RREQ-ACK Cache (as explained in Section – 5.4). To summarize, at the time of RREQ message flooding, RREQ-ACK Cache is updated and at the time of receiving RREP message, the same cache is used for validation.

## 5.7. Preventing the concerned attacks using AODVSEC

With the use of already maintained RREQ-ACK Cache information, each incoming RREP message is first validated and upon success, it would be followed by normal procedure just like AODV. RC, RD and RI, these three attacks are directly prevented through the validation because of the mismatch in timestamp and flag values in the RREQ-ACK Cache. Whereas, in blackhole attack, as explained earlier that the attacker upon receiving any of RREQ message responds with the fake RREP by means of forging destination sequence number and/or hop count fields. In this case, there is a lake of previous RREQ-ACK message from the node replying with RREP message and based on that there would be absence of RREQ-ACK Cache entry and hence the RREP sent by the blackhole would directly be discarded.

## 6. METHODOLOGY OF EVALUATION

In our evaluation, we compare the performances of AODVSEC with the existing AODV and SAODV using Network Simulator – 2.33 (NS-2) [15]-[18]. The details of simulation environment and the performance metrics are given in the following subsections.

### 6.1. Simulation Environment

At the physical and data link layer, we used the IEEE 802.11 with Two Ray Ground radio propagation model. We have considered the traffic of Constant Bit Rate (CBR) data packets over UDP at the transport layer in a rectangular field of 850m x 550m with the total number of 15 nodes forming the ad hoc network.

Table 1 summarizes the complete setup for the simulation.

Table 1. Simulation Parameters.

| Parameter | Value |
| --- | --- |
| Simulator | NS-2.33 |
| Simulation Area | 850m X 550m |
| Propagation model | Two Ray Ground |
| Routing Protocols | AODV, SAODV, AODVSEC |
| Traffic Source | CBR (UDP) |
| # Mobile Nodes | 15 |
| # Connections | 1 |
| Packet Size | 512 byte |
| # Malicious Nodes | 0, 1, 3 |
| Attacks | RC, RD, RI, BH |
| Simulation Time | 500 second |
| Attacking Time | 100s, 200s, 350s |

### 6.2. Performance Metrics

In order to evaluate the performance of the concerned routing protocols, the following five metrics are considered:

1. Packet Delivery Fraction (PDF): This is the ratio of the number of data packets successfully delivered to the destinations to those generated by sources.

2. Normalized Routing Load (NRL): The number of routed packets transmitted per data packet delivered at the destination.

3. Average Throughput (AT): It is the rate of successfully transmitted data packets in a unit time in the network during the simulation.

4. Average End-to-End Delay (AED): It is defined as the average time taken by data packets to propagate from source to destination across the network. This includes all possible delays

International Journal of Computer Networks & Communications (IJCNC) Vol.4, No.4, July 2012

caused by buffering during routing discovery latency, queuing at the interface queue, and retransmission delays at the MAC, propagation and transfer times.

5. Jitter: It is the variation in the time between packets arriving at destination node, caused by network congestion, timing drift, or route changes

## 7. RESULT ANALYSIS

This section presents the simulation results for the following cases:

### 7.1. Normal Case (Without any malicious Activity)

The performance of proposed AODVSEC with respect to five different performance metrics has been compared with that of AODV and SAODV under normal conditions in absence of any malicious activity. Figure 12 and Figure 13 show the graphs of Packet Delivery Fraction, Normalized Routing Load versus Time. From each graph it is evident that performance of AODVSEC, AODV and SAODV are same. For other performance metrics namely AT, AED and Average Jitter also, the performance of trio turned out to the exactly same. (The graphs are not presented here due to space considerations).

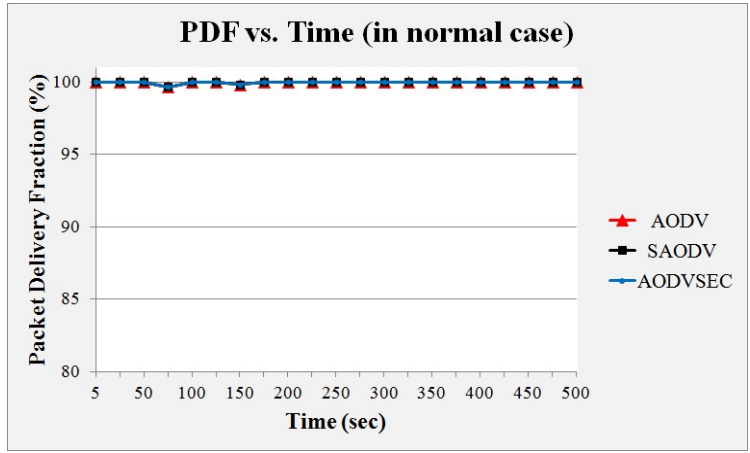

Figure 12. Packet Delivery Fraction vs. Time

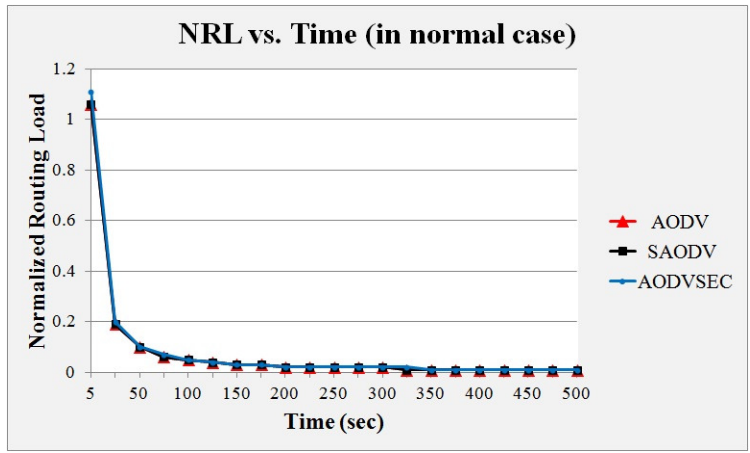

Figure 13. Normalized Routing Load vs. Time

International Journal of Computer Networks & Communications (IJCNC) Vol.4, No.4, July 2012Thus AODVSEC is simpler to implement as compared to SAODV and yields same performance. We simulated these protocols for an additional performance metric, Accumulated Average Processing Time for performing their corresponding functionalities upon receiving specific control message (see Figure 14).

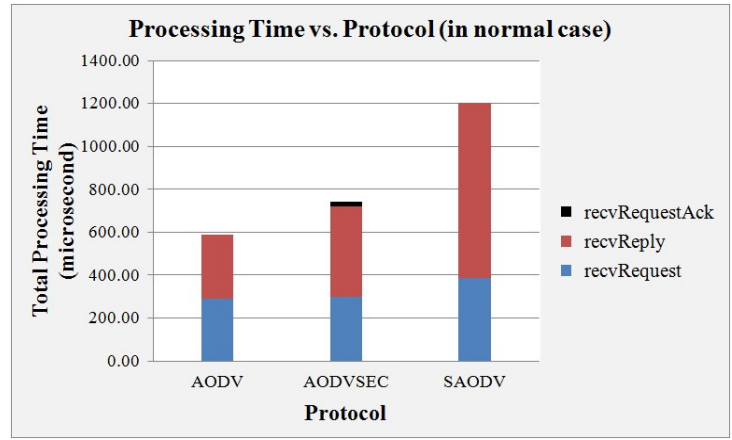

Figure 14. Accumulated Average Processing Time vs. Protocol

From the above Figure 14, it can be concluded that the average processing time of AODVSEC is little higher than that in AODV because of the working for RREQ-ACK Cache during the route discovery process (as explained in section 5). As SAODV applies the cryptographic functions upon receiving any control message to validate the same, it requires higher processing time leading to more energy requirement of the node in comparison to AODV and AODVSEC. Hence in normal condition (without any malicious activity) SAODV consumes more computational power of the participating mobile node whereas AODVSEC does not.

### 7.2. In Presence of Resource Consumption (RC) Attack

This subsection compares the performance evaluation of the protocols in presence of loop forming attack. In this specific attack, data packets are forcefully looped during its delivery to the ultimate destination resulting into data loss (see Figure 15). At the same time, the victim nodes have to participate compulsorily in forwarding duplicated data packets that results into the resource consumption, as well as increase in the routing overhead as well (see Figure 16).

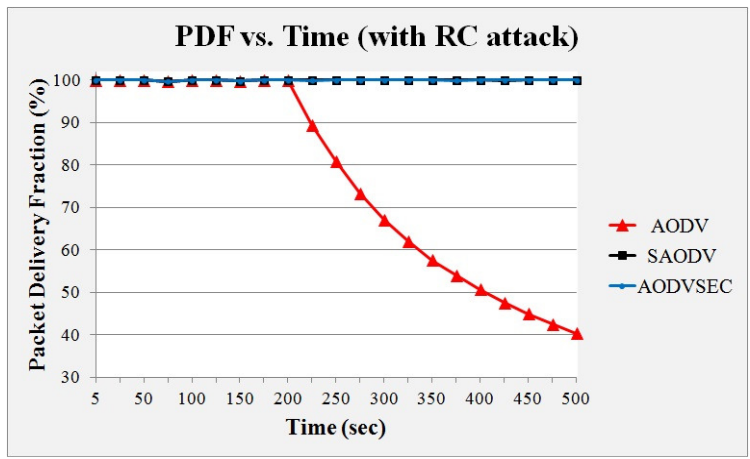

Figure 15. Packet Delivery Fraction vs. Time

203

International Journal of Computer Networks & Communications (IJCNC) Vol.4, No.4, July 2012

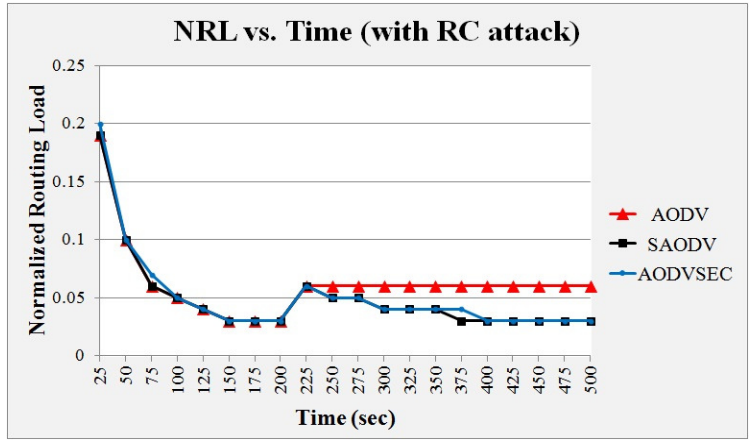

Figure 16. Normalized Routing Load vs. Time

From the above graphs, it can be seen that after the attack at time 200 second, performance of AODV drops whereas AODVSEC and SAODV do not get affected and performance of AODVSEC is at par as that of SAODV.

### 7.3. In Presence of Route Disturb (RD) Attack

This attack periodically tends to either break or disturb the existing route resulting into new route discovery procedure. As a result new route discovery process increases the routing load (see Figure 17), also the ultimate destination may notice increased jitter while receiving the actual data sent by the originator (see Figure 18).

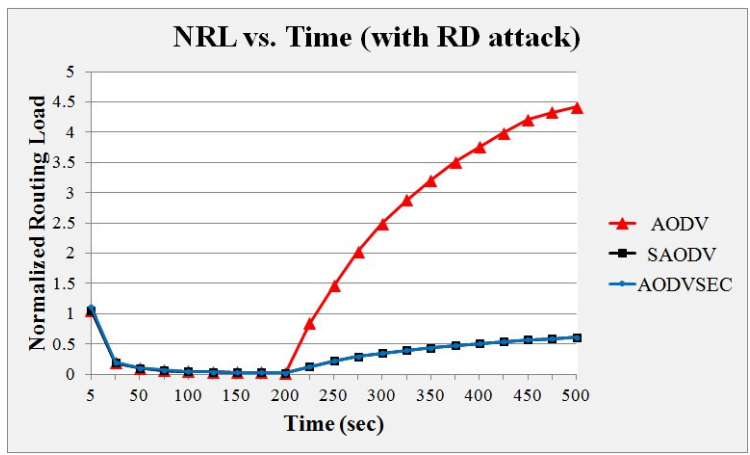

Figure 17. Normalized Routing Load vs. Time

International Journal of Computer Networks & Communications (IJCNC) Vol.4, No.4, July 2012

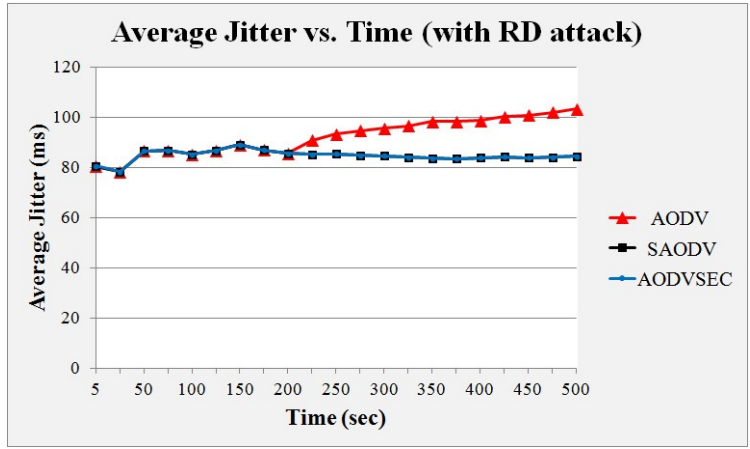

Figure 18. Average Jitter vs. Time

From the above Figures it is observed that – after attacking time (200s), the performance of AODV degrades whereas AODVSEC and SAODV mostly maintain their own normal behavior.

### 7.4. In Presence of Route Invasion (RI) Attack

In this attack, the attacker does not break the route and at the same time the packet delivery to the destination is not affected. Instead it invades itself into the data path so the protocol becomes highly vulnerable to snooping. For showing the effect of this attack we have calculated the total data packets received by the attacker during the whole simulation.

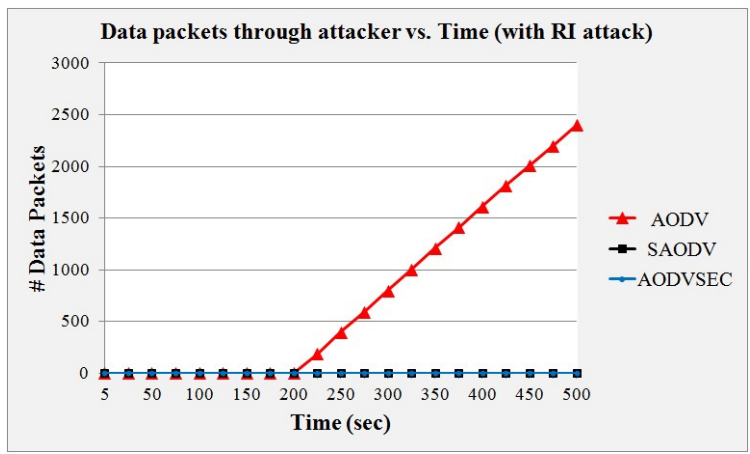

Figure 19. Data Packets though Attacker vs. Time

We can conclude from the above Figure 19 that before the attack (200s), the attacker could not get any data packets in any of the protocol. Once attack is launched, attacker node starts getting data packets under AODV whereas AODVSEC and SAODV remain aloof from the attack.

### 7.5. In Presence of multiple malicious activities

Figure 20 to Figure 23 show results with the three nodes performing the random malicious activities at their own specific time. We have considered a case where all the three attacks are launched separately by three attackers. We did RI attack at 100s followed by RC attack at 200s and then RD attack at 350s.

205

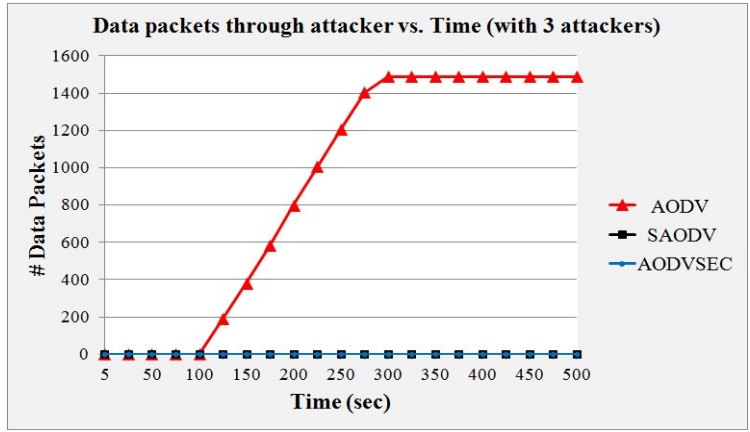

Figure 20. Data Packets though Attacker vs. Time

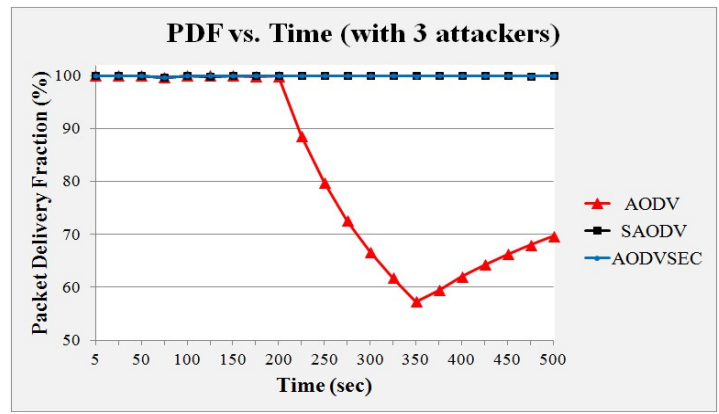

Figure 21. Packet Delivery Fraction vs. Time

Figure 20 shows that, the attacker node started receiving the data packets for AODV, but, after 300s when RC attack has launched, it stopped getting the same as the rest of the data packets were forced in the loop. AODVSEC and SAODV remained unaffected and performed normally. AODV gets worst affected due to RC attack leading to seizable drop in PDF (Figure 21).

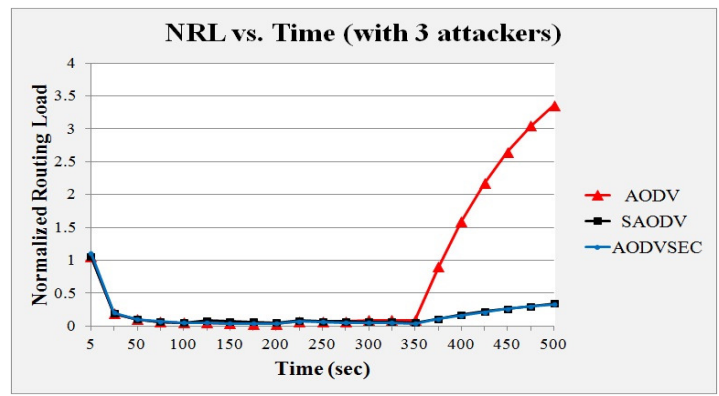

Figure 22. Normalized Routing Load vs. Time

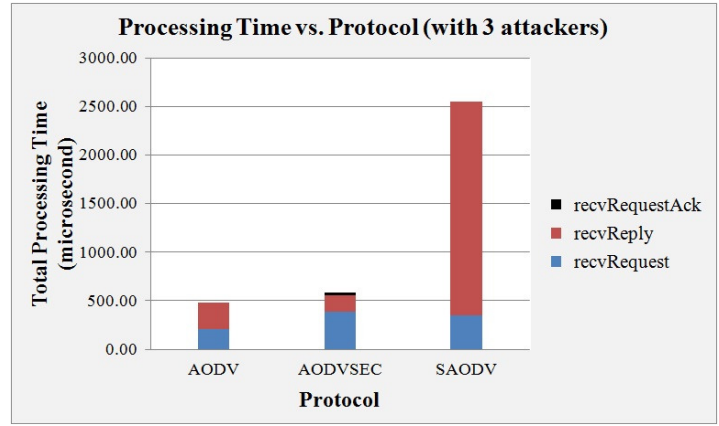

Figure 23. Accumulated Average Processing Time vs. Protocol

Figure 22 shows the performance of NRL in which AODV is comparatively less affected during RC attack than during RD attack (after 350s). In all, AODVSEC and SAODV secure themselves from the attacks. Figure 23 shows the accumulated average processing time of all three protocols in the presences of all three attacks and from this graph it can be concluded that AODVSEC utilizes quite less processing time and ultimately less computational power as compared to SAODV.

### 7.6. In Presence of Blackhole Attack

Figure 24 shows the results with consideration of Blackhole attack. As PDF is the most affected and obvious performance metric we have shown the comparison of all three protocols for this parameter.

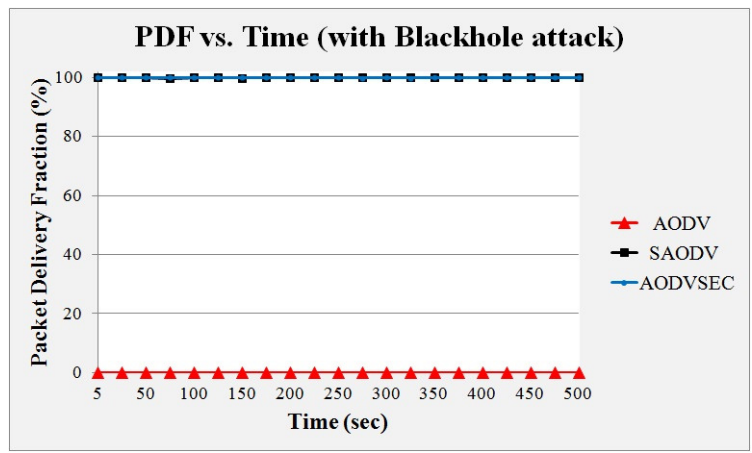

Figure 24. Packet Delivery Fraction vs. Time

It is obvious from the above Figure that AODV fails to transfer any of the sent data packets to the intended destination, whereas AODVSEC and SAODV do not get affected at all by the blackhole attack.

## 8. CONCLUSION & FUTURE WORK

We have modified the fundamental route discovery process of basic AODV in such a way that the protocol can mitigate the effects of active forge attacks viz. Resource Consumption (RC) attack, Route Disturb (RD) attack and Route Invasion (RI) attack through fake RREP message. To provide the security provision, we have not used much complex processing e.g. Cryptography Functions or any logic that further requires the central trusted authority. We have compared AODVSEC with the existing AODV and SADOV protocol. Our results show that the performance of AODVSEC is no less than that of SAODV but the same is achieved with lower processing requirement leading to lots of saving of computational power.

Our future work will focus on security aspects for the attacks that can be launched through forging the RREQ control message. AODVSEC is currently applicable in the scenario where there is presence of the active forge reply based attacks. In all as of now, AODVSEC gives similar results like SAODV and that is with the all important computational power saving although it is attacking condition based on the fake RREP messages.

International Journal of Computer Networks & Communications (IJCNC) Vol.4, No.4, July 2012

## Authors

**Akshai Aggarwal** (MIEEE'1966, SMIEEE'1992, LMIEEE 2011) is working as Vice Chancellor, Gujarat Technological University, Ahmedabad, India. Before joining as the Vice-Chancellor, he was working as the Director of School of Computer Science, University of Windsor, Canada. He worked as Professor and Head of Department of Computer Science at Gujarat University for about 10 years. Before that he was Professor and Head, Department of EE at M.S.University of Baroda. He was Chairman of IEEE India Council for two years. He initiated IEEE activities in Gujarat by starting the first IEEE Student 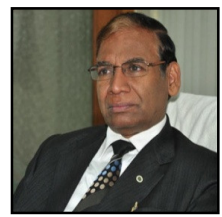 Branch at M.S.University of Baroda. Later he initiated the establishment of the Student Branch at Gujarat University. He was also the founder Chairman of IEEE Gujarat Section, the IEEE Computer Society Chapter and the IEEE Joint Chapter of Industry Applications, Industrial Electronics and Power Electronics. The Section conducted two International Conferences and one national Seminar during his Chairmanship. He graduated with a B.Sc.(EE) from Punjab Engg College and studied at MS University of Baroda for his Master's and Doctoral work.

**Savita Gandhi** (MIEEE' 2003 SMIEEE' 2005) is Professor & Head at the Department of Computer Science, Gujarat University and Joint Director, K.S. School of Business Management, Gujarat University. She is with Gujarat University for about 24 years. Before that she has worked with M.S. University of Baroda, Department of Mathematics for about 10 years. She has been actively associated with IEEE activities at Gujarat Section. She is M.Sc. (Mathematics), Ph.D (Mathematics) and A.A.S.I.(Associate Member of Actuarial Society of 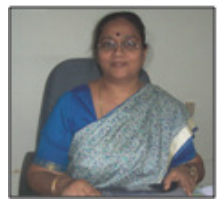 India by the virtue of having completed the "A" group examinations comprising six subjects conducted by     Institute of Actuaries , London). She was awarded Gold Medal for standing first class first securing 93% marks in M.Sc. and several prizes at M.Sc. as well as B.Sc. Examinations for obtaining highest marks.

**Nirbhay Chaubey** (SIEEE' 2002 MIEEE' 2004) pursuing his Ph.D (Computer Science) at Department of Computer Science, Gujarat University, Ahmedabad, India and working as an Assistant Professor of Computer Science at Institute of Science and Technology for Advanced Studies and Research, Vallabh Vidyanagar, Gujarat, India. He has been involved in IEEE activities since 1994. His position held for IEEE Gujarat Section include Executive Secretary (1998-2005), Treasurer (2005-2006), Secretary and Treasurer (2007) and  Treasurer for year 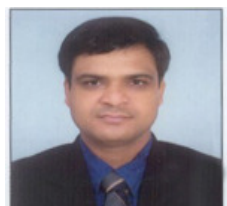 2008 onwards. He graduated from Ranchi University, Ranchi, and Master in Computer Applications from Madurai Kamraj University, Madurai, India.

International Journal of Computer Networks & Communications (IJCNC) Vol.4, No.4, July 2012**Pathik Shah** received his graduation degree for B.E. (Electronics & Communications Engineering) from Gujarat University, Ahmedabad, Gujarat, India in 2010. Currently he is a student at Department of Computer Science, Gujarat University for completing his study of post graduation degree in M.Tech. (Networking & Communications). He opted the domain of Security Concerns in Mobile Ad hoc Networks for his Dissertation work towards the current professional degree.

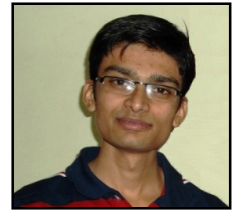

**Madhvi Sadhwani** received her graduation degree for B.E. (Information Technology) from Dharmsinh Desai University, Nadiad, Gujarat, India in 2008. Currently she is a student at Department of Computer Science, Gujarat University for completing her study of post graduation degree in M.Tech. (Networking & Communications). She opted the domain of Security Concerns in Mobile Ad hoc Networks for her Dissertation work towards the current professional degree.

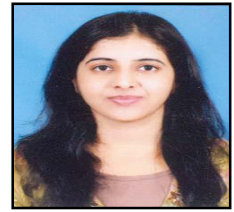